\documentclass{aastex}          
\usepackage{spr-astr-addons}    

\begin{document}
%
\title{Westerlund 1: bound or unbound?}

\shorttitle{Westerlund 1: bound or unbound?}
\shortauthors{Mengel \& Tacconi-Garman}

\author{Mengel, S\altaffilmark{}} 
\and 
\author{Tacconi-Garman, L.E.\altaffilmark{}}
\affil{ESO, Karl-Schwarzschild-Str. 2, 85748 Garching, Germany}
\email{smengel@eso.org} 


\begin{abstract}

The Galactic open cluster Westerlund~1 (Wd 1) represents the ideal local
template for extragalactic young massive star clusters, because it
is currently the only nearby young cluster which reaches a mass of around
$\sim$10$^5$\,M$_\odot$.
The proximity makes spatially resolved studies of its
stellar population feasible, and additionally permits direct comparison 
of its properties with measurements of velocity dispersion and dynamical 
mass for spatially unresolved extragalactic clusters.

Recently, we published the dynamical mass estimate based on spectra of 
four red supergiants. We have now identified six additional stars which
allow a determination of radial velocity from the wavelength covered in
our VLT/ISAAC near-infrared spectra (CO bandhead region near 2.29$\,\mu$m),
improving statistics significantly.
Using a combination of stepping and scanning the slit across the cluster centre, 
we covered an area which included the following suitable spectral types: 
four red supergiants, five yellow 
hypergiants, and one B-type emission line star.

Our measured  velocity dispersion is 9.2\,km/s. Together with the
cluster size of 0.86\,pc, derived from  archival near-infrared SOFI-NTT
images, this yields a dynamical mass of 1.5$\times10^5$M$_\odot$. 
Comparing this to the mass derived via photometry, there is no indication that
the cluster is currently undergoing dissolution.

\end{abstract}


\section{Introduction}
The concept that globular clusters may still be forming today dates back fifteen years
\citep{Hetal92}, and ever since young massive star clusters (YMSCs) have been the
focus of intense studies. First, it needed to be verified that they
actually have the properties which will make them comparable to
(part of) the globular clusters seen today after a Hubble time of evolution.

Once it was clear that, at least regarding mass, size and concentration, they 
could indeed be globular clusters in the making \citep{Whitmore95}, survival times
became an issue. A fraction between 10\% and around 90\% of the
clusters seems to dissolve in each decade of time 
\citep{Fall04, Mengel05, Bastian05, Whitmore06}. At least the
high end of these destruction rates is in conflict with the assumption that
clusters at an age of around 10 Myr, which have typically survived tens of
crossing times, are in virial equilibrium.

However, only clusters which can be spatially resolved into single stars
can be used to verify their dynamical state directly.
The well known  and comparatively numerous Galactic clusters with maximum masses of
around 10$^4$M$_\odot$ are not suitable as templates, because at an age
of around 10 Myr, they are not necessarily expected to be in virial equilibrium,
owing to longer crossing times, and the increased impact of interaction with galactic environment. 
Therefore, Wd 1 turned out to be the best-suited template in the
Milky Way when it was discovered that its mass is much higher than initially
\citep{Westerlund61} thought, around $\times10^5$M$_\odot$ \citep{Clarketal2005}. 
Given its young age of around 5 Myr, there is a considerable probability
to find it in the process of disruption.

In an attempt to determine the dynamical state of Wd1, we conducted high resolution
spectroscopy in the near-infrared. That is the only wavelength range
(because of the high extinction suffered by the cluster) which simultaneously
allows a high signal-to-noise ratio, high spectral resolution, and large
spatial coverage.

\section{Observations and data reduction}

\begin{figure}
\centering
\includegraphics[width=8cm,angle=0]{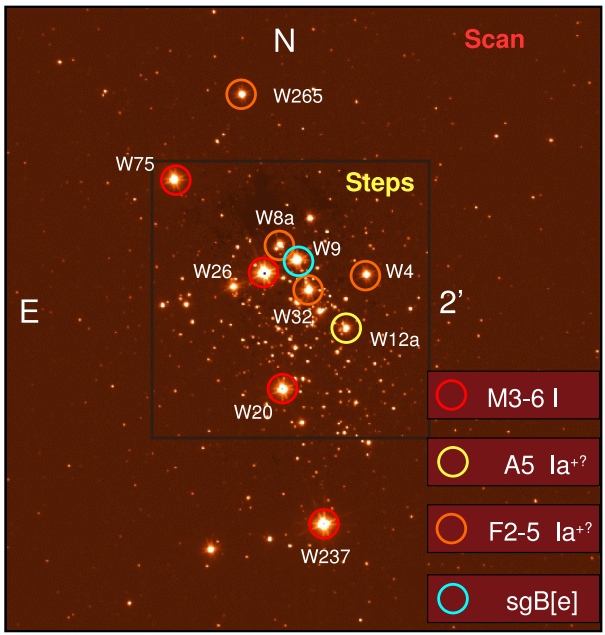}
   \caption{NTT/SOFI near-infrared image of Wd 1, composed from
     narrow band images (2.09$\mu$m and 2.17$\mu$m). The area shown
     was covered by the slit scans (with the slit oriented N-S), and
     the area shown by the black box was covered by steps of 0\farcs3 
     separation. Different stellar types are indicated by different 
     coloured circles, and the stars are identified by their
     \citet{Westerlund87} designation. \label{scanimage}     }
\end{figure}

Observations were conducted on VLT-ANTU on the nights of March 11 and
12, 2006
and consisted of two parts: 1) scanning the slit across a region while integrating,
covering the whole field shown in Figure~\ref{scanimage},
and 2) stepping in steps of 0\farcs3, the slit width, across the smaller
region indicated by the black box in  Figure~\ref{scanimage}.
In both cases, the 2$^\prime$ long slit was oriented in N-S-direction.
The large spatial coverage of the scan was achieved by performing
scans of three different slit positions, offset in N-S-direction by
108$^{\prime\prime}$ each time, which allows for some spatial overlap.

The central wavelength was set to be 2.31\,$\mu$m, and the actual
wavelength range
covered is 2.249 -- 2.373\,$\mu$m at a resolution of R$\approx$9000
(the maximum resolution ISAAC allows).

Reduction of the data was performed using the IRAF
package\footnote{IRAF is distributed by the National Optical Astronomy
Observatories, which are operated by the Association of Universities
for Research in Astronomy, Inc., under cooperative agreement with the
National Science Foundation.}.
It included sky subtraction (a scan across an empty sky
field 10$^\prime$ away was performed with identical integration time
of 300\,s directly after the cluster scan, and sky positions were interleaved
in the stepping observation every 10 minutes), bad pixel correction,
and flatfielding.

Wavelength calibration combined the information of the XeAr lamp
spectra and the night sky lines \citep{Rousselotetal}.
Spatial distortion was easily corrected due to the  many star spectra
along the slit. 

We extracted the individual spectra by averaging all the individual 
frames which included a star of interest, and then extracting from
the combined frame by defining the aperture by hand.

For the correction of the telluric absorption features, we used
the hot, featureless stars covered by the stepped spectra (for example,
W42a and a star without a known designation). This gave a much better
result than using the dedicated solar type telluric standards taken
after the scans, which left some residuals around the MgI line.

Finally, we normalized all spectra which were used for the analysis
by dividing by the continuum level determined from a small window (roughly 50
pixels) just shortward of the CO bandhead.
For the age determination, we used a rough flux calibration as an
additional constraint on the spectral types. Nothing has changed for those
results since \citet{Mengel07}, we refer to that publication for the
details.

\begin{figure*}
\centering
\includegraphics[bb=27 37 591 755, width=10cm,height=17cm,angle=-90,clip]{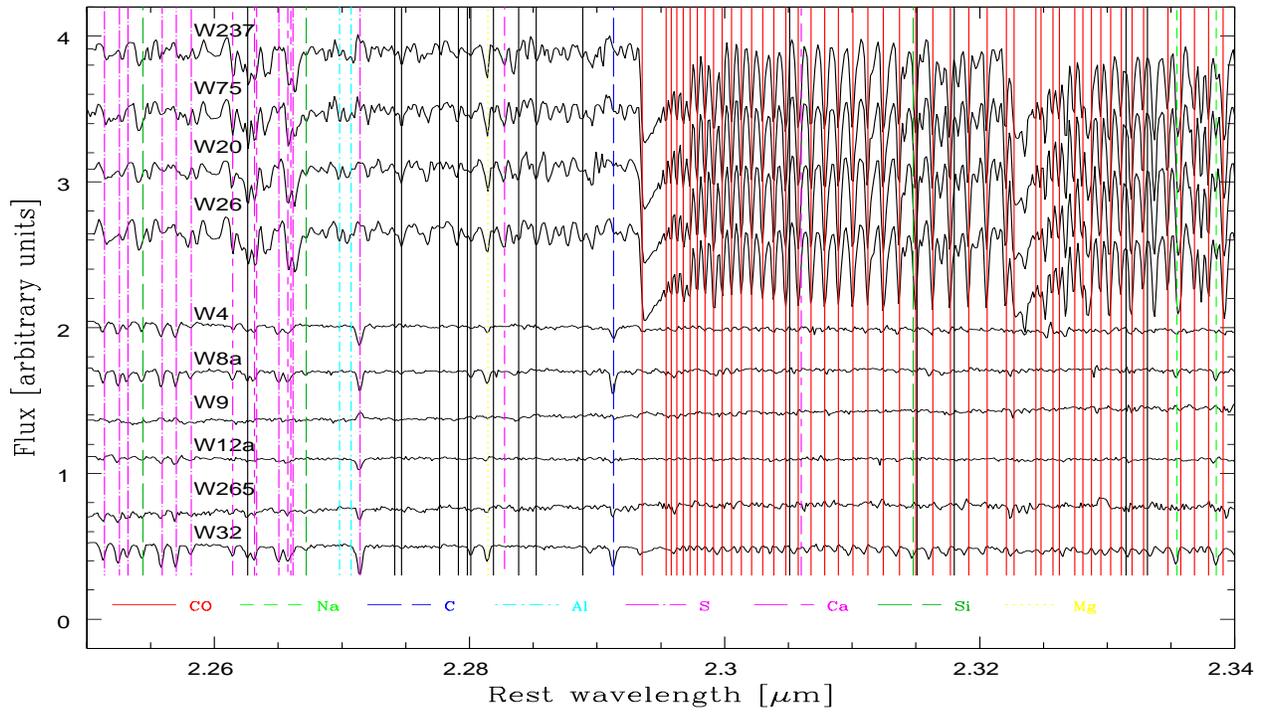}
    \caption{Specta of integrated cluster and the near-infrared
               brigthest individual stars. Atmospheric absorption
             (corrected for, but over-compensation and sub-optimal sky subtraction
              causes a few artefacts at long wavelengths) is shown at
the bottom. Line identifications taken from \citet{Hinkle95}.
       \label{combinedspectra}}
\end{figure*}

The resulting spectra are shown in Fig.~\ref{combinedspectra}.

\section{Analysis and Results}\label{analysisresults}

The four brightest stars are assigned spectral types ranging between M3I and M6I (foreground
objects are ruled out also because of the small spread in radial velocity).

\begin{table*}
\begin{center}
\begin{tabular}{lcccc}
Star          &   Reference star   &  Spectral lines used  & v$_{corr}$ & Spectral Type\\\hline
                &                 &           &       \\\hline\hline
W26             &   W237          &  CO                                               &  -48.7    & M5-6I  \\
W20             &   W237          &  CO                                               &  -49.2    & M5I  \\
W75             &   W237          &  CO                                               &  -59.2    & M4I  \\
W237            &   Arcturus      &  CO                                               &  -61.7    & M3I  \\
W4              &   W32           &  lines below 2.26$\mu$m, S 2.27, Mg 2.28, C2.29   &  -57.8    & F2Ia$^+$  \\
W8a             &   W32           &  lines below 2.26$\mu$m, S 2.27, Mg 2.28, C2.29   &  -41.9    & F5Ia$^+$  \\
W9              &   W32           &  S emission                                       &  -62.4    & sgB[e]  \\
W12a            &   W32           &  lines below 2.26$\mu$m, S 2.27                   &  -61.3    & A5Ia$^+$  \\
W265            &   W32           &  lines below 2.26$\mu$m, S 2.27, C2.29            &  -35.5    & F5Ia$^+$  \\
W32             &   W20, W26      &  Ca 2.26, Mg 2.28                                 &  -57.4    & F5Ia$^+$  \\\hline
\end{tabular}
\caption{Selected properties of the stars used for the measurement of the velocity dispersion:
Designation as in \citet{Westerlund87}, the star which was used as a crosscorrelation reference, 
the spectral lines used for the cross correlation, the heliocentrically corrected radial velocity
(we applied a correction of +27.6km/s), and the spectral types (for the RSGs from \citet{Mengel07},
for the other stars from \citet{Clarketal2005}. See text for an explanation of reference stars and spectral lines.\label{spectratable}}
\end{center}
\end{table*}

We determined the rms radial velocity dispersion of all stars
where this was possible from the wavelength range covered in our
observations, which limited the selection to four red supergiants,
five (assumed) yellow hypergiants (YHGs), and one sgB[e] type 
emission line star.
See Table~\ref{spectratable} for the individual radial velocities, and
also some other properties like designation, spectral type, etc.

A challenge, compared to the determination of velocity dispersion 
from just the RSGs, was the fact that the YHGs and the RSGs have
almost no absorption lines in common at this wavelength range, and we
also (currently) have no access to theoretical high-resolution spectra
in the K-band for the YHGs. Therefore, we used a multi-step-procedure
in an attempt to use a common wavelength reference frame for all our stars
(except W9, which is a special case): First, we cross-correlated the
spectrum of W237 in the region of CO absorption with the rest wavelength
Arcturus spectrum from \citet{Hinkle95}. Cross-correlation of
W237 with the other RGSs around the CO absorption region gave us their
radial velocities relative to W237. Using W20 and W26 as reference stars,
we used the Ca absoption lines between 2.26 and 2.263$\mu$m, and the Mg line
around 2.282$\mu$m to determine the relative velocites of those stars with
respect to W32. And finally, we used five different lines/regions (all
listed in table \ref{spectratable}) for the determination of relative 
velocities between W32 and the other YHGs. The radial velocity of W9 was
determined by manual best fit of the S emission lines.

The resulting velocity dispersion is 9.2$\pm$2.5\,km/s. 
Uncertainties of this value were determined
through Monte Carlo simulations, using the uncertainties in the
individual velocity determinations, but we added 0.5km/s uncertainty
because of the additional uncertainty in matching RSG and YHG velocity reference 
points.

This value is somewhat higher (even though consistent within the uncertainties)
than the value we had obtained previously from the RSGs alone
\citep{Mengel07}. This could have different reasons: The most obvious one 
would be simply statistics. Using only four stars before, statistic uncertainties
are severe. The new value should be somewhat more reliable in that respect,
but here the nature of the additional stars could introduce additional velocity
dispersion. According to \citet{deJager98}, YHGs undergo radial pulsations, which
can cause variations in radial velocity of up to $\approx\pm7$km/s. Literature on
spectral variations of YHGs is scarce, but from what is known, it seems easily 
possible that an increase in velocity dispersion of a few km/s could originate
from this.

The other parameter we require for the determination of the dynamical
mass is the
projected half-light radius.  We used the NIR image we created from
the archival NTT/SOFI
data to obtain a radius which contains half of the total light (in
projection) to be
r$_{hp}$=0\farcs86$\pm$0\farcs14. The uncertainty comes mostly
through assumption of
different locations of the weakly constrained cluster centre.
This value is consistent with the half-mass radius of 1pc (for
stellar masses between 3.4 and 32 M$_\odot$) presented by \citet{Brandner07}.

The resulting dynamical mass, for a cluster in Virial
equilibrium,  is (assuming $\eta = 9.3$, for details
on the formula see
\citet{Mengel02})
\begin{displaymath}
M_{dyn}  =  \eta \sigma^2 r_{hp} / G = 1.5^{\scriptstyle{+0.9}}_{\scriptstyle{-0.7}}\times 10^5\,M_{\odot}
\end{displaymath}
Within the uncertainties, this is consistent with that found by \citet{Clarketal2005}, and somewhat higher 
than the mass determined by \citet{Brandner07}. Both use stellar counts of the
current population (with detection of the main sequence only in the latter
publication) to infer an initial cluster mass using an assumed stellar initial
mass function.

In \citet{Mengel07}, we used the relative numbers of the stellar populations,
together with evolutionary synthesis models,
\citep{Letal99, VazquezLeitherer05} to age date the cluster.
We determine the best fitting age to be 5 Myrs.

We used different methods (with reasonable agreement) to determine L$_V$ for the cluster:
Using the integrated, extinction corrected K-band magnitude and the Starburst99 V-K colour
at 5 Myr, or the integrated V-band magnitude with (substantial) extinction correction,
or the luminosity expected from Starburst99 for the cluster mass of 10$^5$M$_\odot$ expected
from stellar counts. A total luminosity of L$_V \approx 5.8\times10^6$L$_\odot$ leads to
a light-to-mass-ratio of L$_V$/M$_{dyn}\approx40$. 

\begin{figure}
\centering
\includegraphics[bb=1 1 437 355, width=8cm]{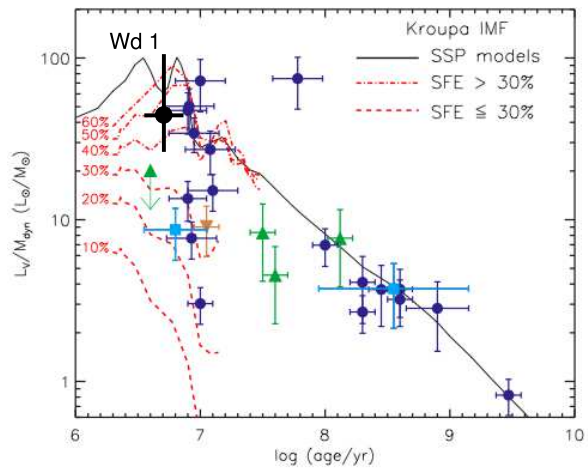}
    \caption{Evolution of L/M$_{dyn}$ for clusters with variable star formation efficiencies
      (modified version of a diagram taken from \citet{GoodwinBastian}).
    Clusters with an effective star formation efficiency of below $\approx$40\% dissolve
    after around 10 Myr. Wd1 (the black cross) is located in a region which would correspond
    to a star formation efficiency of around 45\%, and hence appears stable against (rapid) dissolution.
       \label{GoodwinBastianWd1}}
\end{figure}

Having added our data point in the diagnostic Figure \ref{GoodwinBastianWd1}, taken from
\citet{GoodwinBastian}, we find that Wd 1 is located between the model lines denoting a star formation
efficiency of 40\% and 50\%. Both of these efficiencies, in the N-body models of 
\citet{GoodwinBastian}, lead to the formation of stellar clusters which eventually,
after 10-15 Myr, re-establish virial equilibrium. This means that, based our data,
Wd 1 does not seem to be currently in the process of dissolution.
The potential influence of radial pulsations in the YHG suggests that the velocity
dispersion and hence the virial cluster mass may be overestimated. On the other
hand, mass segregation could have biased our measurements, both the velocity
dispersion, and the cluster size, towards values which are too low.

\section{Conclusions and future work}

    \begin{enumerate}
       \item We identify ten stars which allow a determination of
	 radial velocity in the part of the K-band which we use for
	 our spectral analysis: four RSGs, five YHGs, and one emission line star.
	 The rms velocity dispersion of these
        stars was found to be 9.2$\pm2.5$\,km/s. The resulting
        dynamical mass is $1.5^{\scriptstyle{+0.9}}_{\scriptstyle{-0.7}}\times 10^5\,M_{\odot}$.
       \item There is no indication of cluster dissolution. Comparing L/M to
	 theoretical expectations suggests a star formation efficiency
	 between 40 and 50\%, which is compatible with the cluster re-establishing
	 equilibrium after the phase of gas expulsion.
       \item We cannot rule out systematic or statistical errors in our analysis 
         yet, since the sample of stars, even though larger than before, with
	 just ten stars is still quite small. Furthermore, our measurements
	 may have suffered a bias from either pulsations of the YHGs (velocity
	 dispersion, and hence mass, would be too high), or by not taking into
	 account potential mass segregation (velocity dispersion and size, and
	 hence mass, would be too low). 
    \end{enumerate}

It is rather straightforward to overcome the potential problems mentioned:
For better statistics, we need to have access to more radial velocity 
determinations. We were granted observing time to observe other stellar types
in a different near infrared wavelength region. Furthermore, we will observe
two additional epochs, which will give us an indication of the impact of
radial pulsation in the YHGs. And finally, there are two ways to address
mass segregation: \citet{Brandner07} see some indication of mass segregation
in their SOFI data. Once deeper NIR imaging data, which already exist, 
are analyzed, an even better estimate of this property will be possible. 
Our new spectra will also cover lower mass stars, which will
allow us to check if their velocity dispersion is lower than that of the
lower mass stars. 

All of these future investigation will help settle the question conclusively
if Westerlund 1 will dissolve within the next 10-20 Myr or not - an important
step towards understanding what fraction of the galactic stellar population
originates in star clusters.

\section{Questions and answers}

Mark Gieles: Perhaps there is some evidence for mass segregation: If the high
mass stars sink to the centre, they will speed up, because they move deeper
in the potential. Of course, this argument works only if you can confirm
that your massive stars are also more centrally concentrated.

Sabine Mengel: At the time of the presentation, deep NIR imaging data had not
been published yet, and therefore the main sequence had been undetected
and no information regarding mass segregation was available.
In the meantime, \citet{Brandner07} published their SOFI images, and they
believe that they see some indication for mass segregation, however it
is not quite clear for me yet if this conclusion is unaffected by
statistical and/or observational bias.
It is however believed that the high mass stars segregate because they
lose energy, and that they move more slowly than the low mass stars, 
once they are segregated.

\end{document}